\documentclass[]{spie}
\usepackage[pdftex]{graphicx}
\DeclareGraphicsExtensions{.png,.pdf}

\usepackage[amssymb]{SIunits}
\usepackage[centertags]{amsmath}
\usepackage{nomencl}
\usepackage{amsfonts}
\usepackage{latexsym}
\usepackage{color}
\definecolor{DarkOlive}{rgb}{0.1047,0.2412,0.0064}
\definecolor{FireBrick}{rgb}{0.5812,0.0074,0.0083}
\definecolor{RoyalBlue}{rgb}{0.0236,0.0894,0.6179}
\definecolor{RoyalGreen}{rgb}{0.0236,0.6179,0.0894}
\definecolor{RoyalRed}{rgb}{0.6179,0.0236,0.0894}
\definecolor{LightBlue}{rgb}{0.8544,0.9511,1.0000}
\definecolor{Black}{rgb}{0.0,0.0,0.0}
\definecolor{FuncColor}{rgb}{1.0,0.0,0.0}
\usepackage{mathrsfs}
\usepackage[normalem]{ulem} % for \sout
\usepackage{subfigure}

\newcommand{\Eq}[1]{Equation~(\ref{#1})}
\newcommand{\sq}[1]{\left[ {#1} \right]}
\newcommand{\cu}[1]{\left\{{#1} \right\}}
\newcommand{\ro}[1]{\left( {#1}\right)}
\newcommand{\an}[1]{\left\langle{#1}\right\rangle}
\newcommand{\st}[1]{\left|{#1}\right|}
\newcommand{\du}{\partial}
    \newcommand{\Ref}[1]{Ref.~\citenum{#1}}
    \newcommand{\Fig}[1]{Figure~\ref{#1}}
    \newcommand{\Sec}[1]{Section~\ref{#1}}
\newcommand{\lamD}{\lambda_\mathrm{D}}
\title{Tracking interacting dust: comparison of tracking and state estimation techniques for dusty plasmas} 
\author{Neil P. Oxtoby, Jason F. Ralph, Dmitry Samsonov and C\'eline Durniak
\skiplinehalf
Department of Electrical Engineering and Electronics, 
University of Liverpool, Liverpool, \mbox{L69 3GJ}, United Kingdom
}

\begin{document} 
  \maketitle 

  %%%%%%%%%%%%%%%%%%%%%%%%%%%%%%%%%%%%%%%%%%%%%%%%%%%%%%%%%%%%% 
  \begin{abstract}
    When tracking a target particle that is interacting with nearest neighbors in 
		a known way, positional data of the neighbors can be used to improve the state 
		estimate.  
		Effects of the accuracy of such positional data on the target track accuracy are 
		investigated in this paper, in the context of dusty plasmas.  
		In kinematic simulations, notable improvement in the target track accuracy was 
		found when including all nearest neighbors in the state estimation filter and 
		tracking algorithm, whereas the track accuracy was not significantly improved 
		by higher-accuracy measurement techniques.  
		The state estimation algorithm, involving an extended Kalman filter, was shown to 
		either remove or significantly reduce errors due to ``pixel-locking''.  
		{It is concluded that the significant extra complexity and computational 
		expense to achieve these relatively small improvements are likely to be 
		unwarranted for many situations.}{For the purposes of determining the precise 
		particle locations, it is concluded that the simplified state estimation 
		algorithm can be a viable alternative to using more computationally-intensive 
		measurement techniques.}
  \end{abstract}

    %%%%%%%%%%%%%%%%%%%%%%%%%%%%%%%%%%%%%%%%%%%%%%%%%%%%%%%%%%%%%
    \section{INTRODUCTION}\label{sec:intro}
      Dusty, or complex, plasmas consist of a low-density neutral/ion/electron plasma containing 
      suspended ``dust'' -- negatively-charged macroscopic particles.  Dusty plasmas occur in space 
  	  as well as in various terrestrial discharges ranging from lightning to industrial 
  	  applications.\cite{Merlino2004}
  	  In more controlled environments, dusty plasmas offer a unique testbed for exploring 
  	  very complex and fascinating physical processes on a kinematic level.  
  	  In all cases, precise determination of properties of the dusty plasma is a top 
  	  priority.  {In particular, the industrial applications of dusty plasmas could benefit 
  	  from being able to determine, and subsequently control, physical properties 
  	  of a dusty plasma.}

\begin{figure}[ht]
        {\centering
          {\includegraphics[width=0.95\columnwidth]{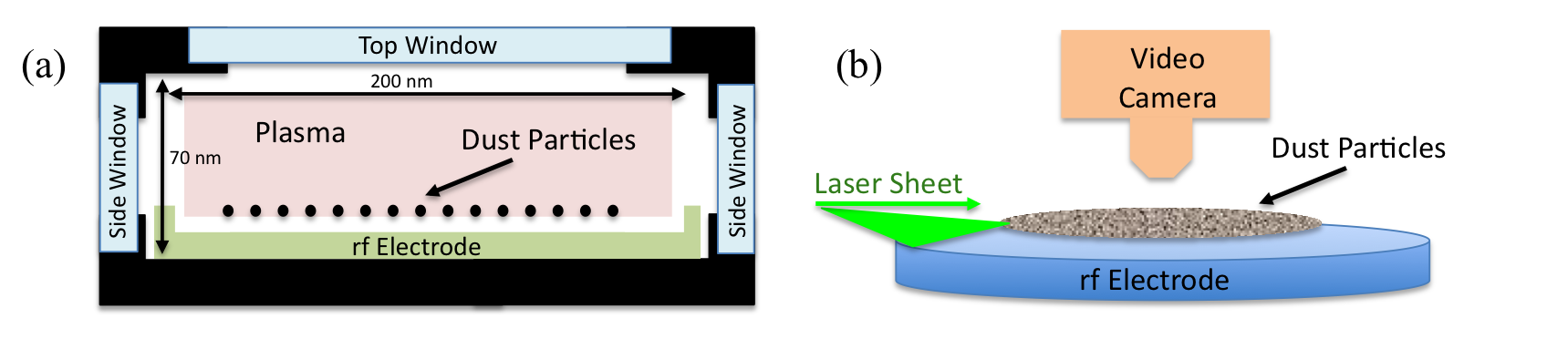}}
          \caption{\label{fig:setup}
            Schematic of the dusty plasma experimental setup.\cite{Durniak2010}
            The dusty plasma is contained wholly within a chamber as shown in (a), 
            with the dust illuminated by a laser sheet and imaged from above by a 
            digital camera as shown in (b).
            }
        }
      \end{figure}

  	  Many properties of a dusty plasma can be inferred from the behavior of the 
  	  dust particles, which can be non-invasively observed over time via laser illumination 
  	  and a digital camera, as in \Fig{fig:setup} which is used in experiments 
  	  on Mach cones\cite{Samsonov1999} and shock-waves\cite{Samsonov2004}.  
  	  The particles acquire a charge due to collisions with ions and electrons in the plasma, 
  	  causing them to interact with each other and the plasma, with often fascinating results.  
  	  For example, the dust can form crystalline structures or behave in a liquid-like phase 
  	  (or exhibit phase transitions\cite{Samsonov2004}), 
  	  on timescales amenable to observation at the individual particle level.  
  	  The precision of one's knowledge of the dust behavior is 
  	  crucial to the reliability of any conclusions made about physical properties of the 
  	  system\cite{hadziavdic2006}.  There are two ways to improve the precision of determining 
  	  the dynamics of individual particles: more sophisticated measurements, and information 
  	  processing.  The performance of each are considered in this work.

  	  Determining the tracks of multiple, interacting targets is a non-trivial task.  
  	  Dust particles are typically observed using a video camera, their locations 
  	  revealed using computer-based image-processing techniques, and their tracks 
  	  (connected positions in time) formed using track-association algorithms.  
  	  Such techniques and algorithms are available in varying levels of sophistication, 
  	  with commensurate levels of accuracy, precision, resource consumption, etc.  
  	  That is, more sophisticated techniques typically provide better precision and 
  	  consume more resources such as time and computer memory.  This article considers 
  	  tracking performance and, specifically, a trade-off between resources and 
  	  accuracy/precision.  Motivation for such considerations comes from 
  	  {the ultimate goal of implementing closed-loop control for such experiments, for which 
  	  the tracking procedure must be automated.}  
  	  
  	  Previous work on tracking individual dusty plasma particles has been somewhat 
  	  unsophisticated.  A common technique is particle tracking velocimetry (PTV), 
  	  where the distance and direction of particle travel is obtained from consecutive 
  	  images and used to aid in linking particle locations 
  	  into tracks.\cite{Samsonov2000,BoesseAiSR04}  This process can be automated, but more can be done.  
  	  A higher level in sophistication for particle tracking is offered by the 
  	  Kalman filter,\cite{BarShalom,JFRopaedia} which combines a prediction model 
  	  with measurements to obtain a recursive state estimation algorithm that can be 
  	  optimal in the minimum-mean-square-error sense for linear dynamics.  
  	  A simplified Kalman filter has been used on simulated data in the context of 
  	  a dusty plasma crystal,\cite{hadziavdic2006} where it was shown to perform very 
  	  well in many situations.  
  	  Slightly more sophisticated still is the state-estimation algorithm known as the 
  	  Extended Kalman filter (EKF)\cite{BarShalom,JFRopaedia}, which is considered in this 
  	  work and will be introduced in \Sec{sec:filter}.
  	  
    %%%%%%%%%%%%%%%%%%%%%%%%%%%%%%%%%%%%%%%%%%%%%%%%%%%%%%%%%%%%%
    \section{SIMULATED DYNAMICS}\label{sec:dynamics}
      Simulating experiments affords this study two primary benefits.  It allows for 
      a larger exploration of ``parameter space'' than might be achievable 
      experimentally in a practical period of time.  In addition, it allows for the 
      \emph{accuracy} of the algorithms to be quantified because the ``true'' 
      particle locations are known.  In the future these algorithms will be 
      developed for and tested on experimental data, so it is important to be as 
      realistic as possible at this early stage.  To this end, the parameters 
      and images of recent dusty plasma experiments\cite{Samsonov1999,Samsonov2004} 
      are used as a guide for this work.
      
      Three characteristic length scales of importance for dusty plasmas are the particle 
      radius $R$, the Debye screening length $\lamD$ and the inter-particle separation 
      $r$.  For the case where $R \ll \lamD < r$, the dust particle interactions can be 
      described well\cite{Resendes1998} by a point-charge approximation.  
  	  Each particle of charge $Q_\mathrm{d}$ experiences forces from other particles 
  	  directly, as well as indirectly via the plasma (such as drag forces due to 
  	  {ion motion}{motion of neutrals} within the plasma).  Under the point-charge approximation, the 
      effective potential experienced by one dust particle (labeled $\jmath$) due to 
      another (labeled $k$) consists of a screened (exponentially decreasing), repulsive 
      Coulomb interaction of the Debye-H\"uckel/Yukawa form:\cite{Konopka2000}
      \begin{eqnarray}
      	\phi_{\jmath,k}(\vec{r}_{\jmath,k}) &= & 
      	  k_0 Q_\mathrm{d} \frac{e^{-r/\lamD}}{r} {- \frac{1}{2\lamD}}{} \hat{r}_{\jmath,k} .
      	\label{eq:phi}
      \end{eqnarray}
      Here $k_0=1/(4\pi\varepsilon_0)$ is Coulomb's constant, the plasma screening distance 
      is the Debye length ($\lamD$), and the particles are separated by 
      $\vec{r}_{\jmath,k} = r \hat{r}_{\jmath,k}$ (note that this separation is time-varying).  
      Assuming a circular two-dimensional geometry (no angular variation), and 
      defining $\tilde{r}\equiv r/\lamD$, yields the effective force between two 
      interacting dust particles (the gradient of the potential) to be
      \begin{equation}
      	F(\vec{r}_{\jmath,k}) = 
      	  \frac{k_0 Q_\mathrm{d}^2}{\lamD^2} 
      	  e^{-\tilde{r}} \ro{\tilde{r}^{-2} + \tilde{r}^{-1} {- \frac{1}{2}}{}} \hat{r}_{\jmath,k} .
      	\label{eq:force}
      \end{equation}
      {which is zero at a particle separation of $r = (1+\sqrt{3})\lamD \equiv r_0$.}{}
      {For $r>r_0$, two simulated dust particles attract each other weakly, and for $r<r_0$ 
      they repel strongly.
      When used as the initial separation between simulated dust particles, small (thermal, etc.) 
      deviations can be expected about this minimum, resulting in a crystal-like structure 
      for which an EKF is used for the state estimation procedure 
      (see Section~\ref{sec:filter}).}
      {When the combined effect of surrounding particles and a 
      global confinement potential are included, dusty plasma particles tend to align themselves 
      in a hexagonal lattice\cite{Durniak2010}, about which small (thermal, etc.) deviations can 
      be expected after a transient period.  This situation is considered here.}
      
      The total force on the $\jmath$th particle is then the sum over $k$ for the 
      two-particle interactions (\ref{eq:force}) and the particle dynamics are determined 
      using a fourth-order Runge-Kutta numerical integration of the equations of motion 
      for position $q(t)$ and velocity $v(t)$:
  	  \begin{align}
  	    \frac{dq(t)}{dt} &= v(t) + \sqrt{2D} \dot{W}(t), \\
  	    \frac{dv(t)}{dt} &= \frac{F}{m} .
  	  \end{align}
  	  Brownian motion of the dust particles due to collisions with plasma ions 
  	  has been included in a standard way: a zero-mean delta-correlated stationary Gaussian 
  	  process $\dot{W}(t) = dW(t)/dt$, which satisfies
  	  \begin{align}
  	    \an{\dot{W}(t)} &= 0 ,\label{eq:dW:mean} \\
  	    \an{\dot{W}(t)\dot{W}(t^\prime)} &= \delta(t-t^\prime). \label{eq:dW:var}
  	  \end{align}
  	  The magnitude of a particle's position fluctuations induced by Brownian motion is related to 
  	  the local temperature $T$ by the Einstein relation: 
  	  $D=k_\mathrm{B}T/\gamma m$, where 
  	  $k_\mathrm{B}$ is Boltzmann's constant,
  	  $\gamma$ is a particle damping coefficient, 
  	  and $m$ is the particle mass ($\approx 5\times10^{-13}$ kg in \Ref{Durniak2010}).  
      
      \subsection{Images} % (fold)
      \label{sub:images}
        In a dusty plasma experiment, observations consist of a time sequence of images taken 
        (typically) at regular intervals of $\Delta{t}$.  Thus, when developing a state estimation 
        and tracking algorithm using simulated data, it is beneficial to reproduce images that 
        approximate those obtained experimentally.  
        
\begin{figure}[ht]
          {\centering
            \subfigure[~\label{fig:experiment}]{\includegraphics[width=0.24\columnwidth]{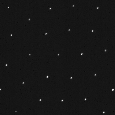}} 
            \subfigure[~\label{fig:simulation}]{\includegraphics[width=0.24\columnwidth,angle=90]{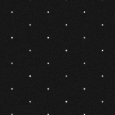}} %\\
            \subfigure[~\label{fig:singlecell}]{\includegraphics[width=0.24\columnwidth,angle=90]{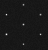}} 
            \subfigure[~\label{fig:defocussed}]{\includegraphics[width=0.24\columnwidth,angle=90]{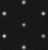}}
          \caption{\label{fig:imageGen2}Zoomed versions of 
                  \subref{fig:experiment} a captured image from experiment, 
                  \subref{fig:simulation} a simulated image without camera defocussing, 
                  \subref{fig:singlecell} a simulated single hexagonal lattice ``cell'' consisting of a central target 
                      particle surrounded by the six nearest neighbors.  
                  \subref{fig:defocussed} a simulated image \emph{with} simulated camera defocussing to 
                      increase pixel-resolution 
                      and reduce/remove pixel-locking errors.
                  Images from experiments and simulations compare very well.}
              }
        \end{figure}

        Experimental images from {}{Refs.~\citenum{Samsonov2004}, \citenum{Feng2007} and 
        \citenum{RalphSPIE09}} provide the benchmark for the images generated from simulations in this 
        work.  The raw images {}{from the experiments reported in Ref.~\citenum{RalphSPIE09}} compare 
        very well to simulated images, as shown in Figures \ref{fig:experiment} and \ref{fig:simulation}.  

      \subsection{Nearest-Neighbor Approximation}\label{sec:NN}
\begin{figure}[ht]
          \centering
            \includegraphics[width=0.5\textwidth]{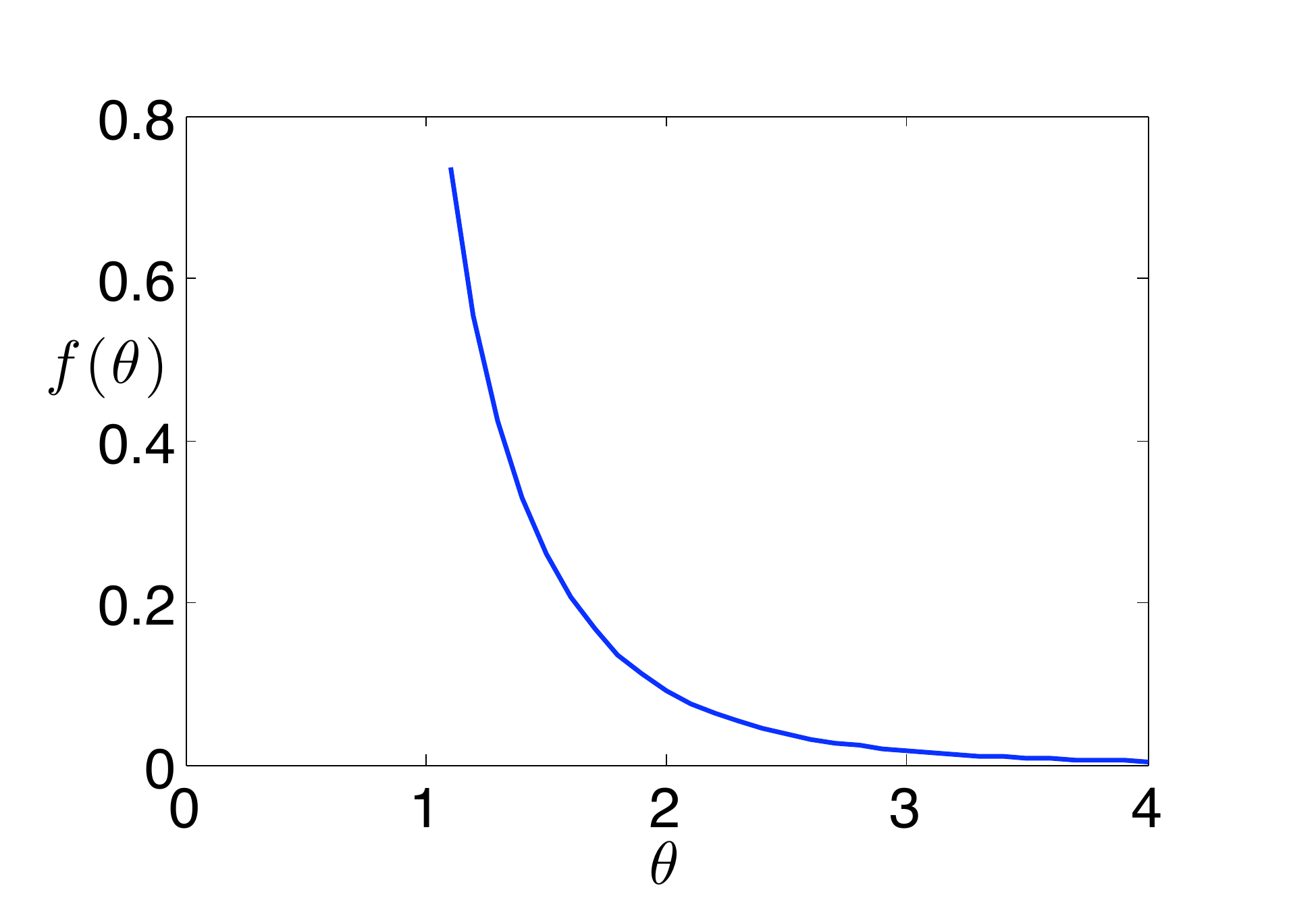}
          \caption{\label{fig:RelError}
            Plot of \Eq{eq:RelError} showing the effect of considering the next nearest neighbors (NNNs) 
            in the tracking algorithm filter.  The uncertainty in the predicted force on the tracked central particle 
            is insignificant ($f(\theta) < 10\%$, say) for $\theta >\sim 2$, which is always true for crystal-like 
            structures of dusty plasmas where $r_\mathrm{NNN} \approx 2 r_\mathrm{NN}$.  If the NNNs move 
            much closer than this, say $\theta < 1.5$, then the influence of the NNNs may need to be considered 
            in the tracking and state estimation algorithm.}
        \end{figure}

        Given the exponential suppression of the Yukawa force in \Eq{eq:force}, it is 
        reasonable to consider the relative effect of the Nearest-Neighbours (NNs) and the 
        Next-Nearest-Neighbors (NNNs) on a ``target'' particle.  
        For a crystal-like state where $r>\lamD$, the exponential suppression factor 
        suggests that the NNNs have relatively little influence on a target's dynamics.  
        We will now show that this is true by analyzing propagation of uncertainties.  
        Uncertainties in the two-particle separation $r$ propagate into $F(r)$ 
        as $\delta F(r) = \st{(\du F / \du r)} \delta r$, where $\du F/ \du r$ calculated from 
        \Eq{eq:force} is
        \begin{equation}
          \frac{\du F}{\du r}(\tilde{r}) = 
          {\frac{k_0Q_\mathrm{d}^2}{\lamD^3} e^{-\tilde{r}}
          \tilde{r}^{-3}\ro{2 + 2\tilde{r} + \tilde{r}^2 - \frac{1}{2}\tilde{r}^3}}
          {\frac{k_0Q_\mathrm{d}^2}{\lamD^3} e^{-\tilde{r}}
          \tilde{r}^{-3}\ro{2 + 2\tilde{r} + \tilde{r}^2}}
          \label{eq:dFdr}
        \end{equation}
        and the uncertainty in the particle separation will typically be 
        $\delta r \leq \sqrt{2} \delta q$, where $\delta q$ is the largest 
        measurement uncertainty in either dimension.  
        That is, the average error in locating a particle's center. 
        In this work, $\delta q \sim 0.1$ pixels. 
        Consider now the ratio between uncertainties in the two-particle forces due to NNs and NNNs: 
        $\delta F (r_\mathrm{NNN}) / \delta F (r_\mathrm{NN})$.  If this ratio is sufficiently small, 
        then we can safely omit the NNNs from the tracking algorithm.  
        Defining $\theta \equiv r_\mathrm{NNN}/r_\mathrm{NN} > 1$ and working in units of 
        $\tilde{r}_\mathrm{NN} \equiv r_\mathrm{NN}/\lamD$, we find that
        \begin{equation}
          f(\theta) \equiv
          \frac{\delta F (r_\mathrm{NNN})}{\delta F (r_\mathrm{NN})} 
            = 
            {\frac{e^{-(\theta-1)}}{9} \st{1 - 2\theta^{-1} - 4\theta^{-2} - 4\theta^{-3}}}
            {\frac{e^{-(\theta-1)}}{5} \ro{2\theta^{-3} + 2\theta^{-2} + \theta^{-1}}} ,
            \label{eq:RelError}
        \end{equation}
        which is approaching insignificance ($f(\theta) < 10\%$, say) for $\theta$ larger than 
        approximately 2, as shown in \Fig{fig:RelError}.  
        Thus, for crystal-like states where $r_\mathrm{NNN} \approx 2 r_\mathrm{NN}$, the tracking 
        algorithm used need only consider the NNs and not the NNNs.  This is a great reduction in 
        complexity.

      \subsection{Simulating a single hexagonal lattice cell} % (fold)
      \label{sub:simulation}
        As shown in the preceding section, for a dusty plasma in a crystal-like phase, a target particle 
        is only significantly affected by its NNs.  Therefore, for the purposes of 
        ascertaining (and quantifying) how the knowledge of surrounding particles can benefit the tracking 
        process, this work focusses on a single cell of a hexagonal lattice, as seen in the 
        experimental and theoretical work of Refs.~\citenum{Samsonov1999}, \citenum{Samsonov2004}, 
        \citenum{Samsonov2000}, \citenum{Durniak2010}, and \citenum{Feng2007}, for example.
        A small set of seven particles was simulated, representing a single cell of NNs surrounding a 
        target particle in a hexagonal lattice as can be seen in Figures~\ref{fig:singlecell} and 
        \ref{fig:defocussed}.  
        The seven particles are initially positioned at an equidistant separation 
        near potential minima (total potential: Yukawa plus confining potential) 
        and interact via the Yukawa-type of potential (\ref{eq:phi}) as described in \Sec{sec:dynamics}.  
        The shallow, parabolic, global confining potential is used to mimic the combined effect of many 
        other surrounding particles and an experimental confining potential.\cite{Durniak2010}  
        These assumptions produce a reasonable model for a small subset of a larger two-dimensional 
        dusty plasma in a crystal-like state.

      % subsection simulation (end)

    %%%%%%%%%%%%%%%%%%%%%%%%%%%%%%%%%%%%%%%%%%%%%%%%%%%%%%%%%%%%%
    \section{STATE ESTIMATION AND TRACKING}\label{sec:filter}
      \subsection{Measurement: image-processing}
        Once images are obtained from a dusty plasma experiment (real or simulated), they are 
        processed to determine the centroids (sub-pixel resolution location) of each particle.  
        Herein, these centroids will be referred to as the particle \emph{measurements}.  
        The centroid location is achieved using one of a number of techniques described in 
        \Ref{Ivanov2007}.  
        The simplest of these is the Threshold Method (TM) where contiguous pixels above a certain 
        threshold are considered to be a particle, with the pixel center-of-mass used for the 
        centroid.  
        The Moment Method (MM) works in a similar way to the TM, but calculates a center-of-mass 
        weighted by the particle's pixel-intensities.  
        A third technique known as the Moment Method with Filter (MMF) involves applying a 
        spatial bandpass filter (typically Gaussian) for smoothing and background subtraction 
        prior to applying the MM to find the weighted center-of-mass.  
        In typical order of increasing precision, these techniques are ranked TM, MM, MMF.  
        There exist more sophisticated (and often more precise) measurement techniques, 
        such as the Linear Quadratic Kernel Method\cite{Ivanov2007}, for which 
        the increased computational expense cannot necessarily be justified 
        for all applications, particularly if combining the measurements with a state-estimation 
        algorithm that can improve the precision of the estimated positions for less computational 
        expense.  
        
        Another option for improving the precision of dusty plasma particle measurements involves 
        a hardware modification: defocussing the camera lens to obtain a ``smeared'' image to which 
        the simpler (and faster) TM and MM methods can be applied.  This approach (used here) 
        helps to avoid errors due to ``pixel-locking''\cite{Feng2007,Ivanov2007} where certain 
        sub-pixel locations are preferred due to low pixel-resolution caused when a particle 
        only illuminates a small number of pixels.  
        The particles in \Fig{fig:singlecell} are of the order of 3 pixels across, whereas 
        those in \Fig{fig:defocussed} are approximately twice as large.  
        The level of smearing can be modified to obtain a desired measurement precision, 
        but it should be sufficiently moderate so as to not affect the measurement-track association 
        process discussed in \Sec{sec:measTrack}.

  	  \subsection{Measurement-track association}\label{sec:measTrack}
  	    Each measurement must be either associated to a track, dismissed as a false alarm, or saved 
  	    in memory as a possible new track.  Along with the task of allowing for missed detections, 
  	    this decision process is known as measurement-track association.  
  	    In multi-target situations this can be a highly nontrivial task, especially when two or more 
  	    targets are close to one another, for example.  
  	    For simple situations where the targets remain spatially well-separated between subsequent 
  	    measurements (Euclidean separation is greater than the measurement and prediction noise), 
  	    a simple radar-like technique can be used to locate the measurement nearest to each track.  
  	    This is the case for crystal-like dusty plasma structures close to equilibrium. 
  	    {This rather simple scenario provides a benchmark for measuring the performance of 
  	    more sophisticated measurement-track association algorithms that are necessary for highly 
  	    nonlinear dynamics such as Mach cones\cite{Samsonov1999} and shock-waves\cite{Samsonov2004}.}
  	    
  	  \subsection{The Extended Kalman Filter}
  	    The state estimation for the particle tracks uses a standard discrete-time Extended Kalman 
  	    Filter (EKF)\cite{BarShalom,JFRopaedia}. The state of each particle being tracked is represented 
  	    by six variables -- two positions $q_\mathrm{x}$ and $q_\mathrm{y}$, two velocities 
  	    $v_\mathrm{x}$ and $v_\mathrm{y}$, and two accelerations $a_\mathrm{x}$ and $a_\mathrm{y}$ -- 
  	    at each time step $k=0,1,2\ldots$. The state is contained in a six-element vector: 
  	    $\hat{x}(k)=(q_\mathrm{x}, v_\mathrm{x}, a_\mathrm{x}, q_\mathrm{y}, v_\mathrm{y}, a_\mathrm{y})^T$ 
  	    and the measurement for each particle is represented by a $2\times 6$ measurement matrix $H$:
        \begin{equation}\label{MeasurementMatrix}
          H=\left(\begin{array}{cccccc}
          	1 & 0 & 0 & 0 & 0 & 0 \\
          	0 & 0 & 0 & 1 & 0 & 0 
          \end{array}\right)
        \end{equation}
        so that the expected measurement for each track, given the previous measurements up to and 
        including $k-1$, is given by
        \begin{equation}\label{Measure}
          z(k|k-1)=H\cdot \hat{x}(k)=\left(\begin{array}{c}
          q_\mathrm{x} (k) \\ q_\mathrm{y} (k) 
          \end{array}\right) .
        \end{equation}
        The expected and actual measurement ($z(k)$) at time step $k$ are combined to form the 
        innovation (the difference between the expected positions and the measured positions) 
        and added to the state vector as in the standard Kalman filter,
        \begin{equation}\label{Update}
          \hat{x}(k_{+}) = \hat{x}(k)+K(k)\cdot (z(k) - z(k|k)) ,
        \end{equation}
        where $\hat{x}(k_{+})$ indicates the updated state vector, the Kalman gain is given by 
        $K(k) = S_{\hat{x}}(k)\cdot H^T\cdot (H\cdot S_{\hat{x}}(k)\cdot H^T+R(k))^{-1}$, $S_{\hat{x}}(k)$ is the estimated 
        covariance matrix for the track states, and $R(k)$ is the expected covariance matrix for 
        the measurement process being used. 

        After each time step, the particle state vector is predicted forward to the next time 
        step using a discretization of the dynamical equations used in the simulation and given 
        in \Sec{sec:dynamics}
        \begin{equation}\label{nonlinStateTrans}
          {\hat{x}}(k+1) = f(\hat{x}(k_{+}))=\left(\begin{array}{c}
          q_\mathrm{x} (k_{+}) +(\Delta t)v_\mathrm{x}(k_{+})+\frac{(\Delta t)^2 F_\mathrm{x}(\hat{x}(k_{+}))}{2m}\\ 
          v_\mathrm{x}(k_{+})+\frac{(\Delta t)F_\mathrm{x}(\hat{x}(k_{+}))}{m}\\ 
          a_\mathrm{x}(k_{+})\\ 
          q_\mathrm{y} (k_{+}) +(\Delta t)v_\mathrm{y}(k_{+})+\frac{(\Delta t)^2 F_\mathrm{y}(\hat{x}(k_{+}))}{2m}\\ 
          v_\mathrm{y}(k_{+})+\frac{(\Delta t)F_\mathrm{y}(\hat{x}(k_{+}))}{m}\\ 
          a_\mathrm{y}(k_{+})\\ 
          \end{array}\right) ,
        \end{equation}
        where $\Delta t$ is the time step. The corresponding equation for the predicted covariance is
        \begin{equation}\label{nonlinErrorTransition}
          S_{\hat{x}}(k+1) = f'(k)\cdot S_{\hat{x}}(k_{+})\cdot f'(k)^T + Q(k) ,
        \end{equation}
        where $Q(k)$ is the process noise covariance, and
        \begin{equation}\label{nonlinErrorTransition2}
          f'(k)=\left.\frac{\partial f}{\partial \hat{x}}\right |_{\hat{x} = \hat{x}(k_{+})} .
        \end{equation}
        This constitutes a piecewise linearization of the nonlinear particle dynamics represented 
        by the simulation. This linear approximation should be accurate as long as the errors 
        in the state estimates are small compared to the nonlinear nature of the underlying 
        potential function. For the cases considered here, with the simulated dust particles 
        in a crystal-like state, the relative movement of the particles around their equilibrium 
        positions will be small enough for this approximation to hold. In more dynamic dusty plasmas, 
        checks would be required to ensure that the estimated errors were not exceeding the ranges 
        required for the perturbative expansion of the inter-particle forces about the current state estimates.
  	    
  	  \subsection{Process noise tuning}\label{sec:PNtune}
\begin{figure}[htp]
    	    {\centering
    	      \includegraphics[width=0.7\textwidth]{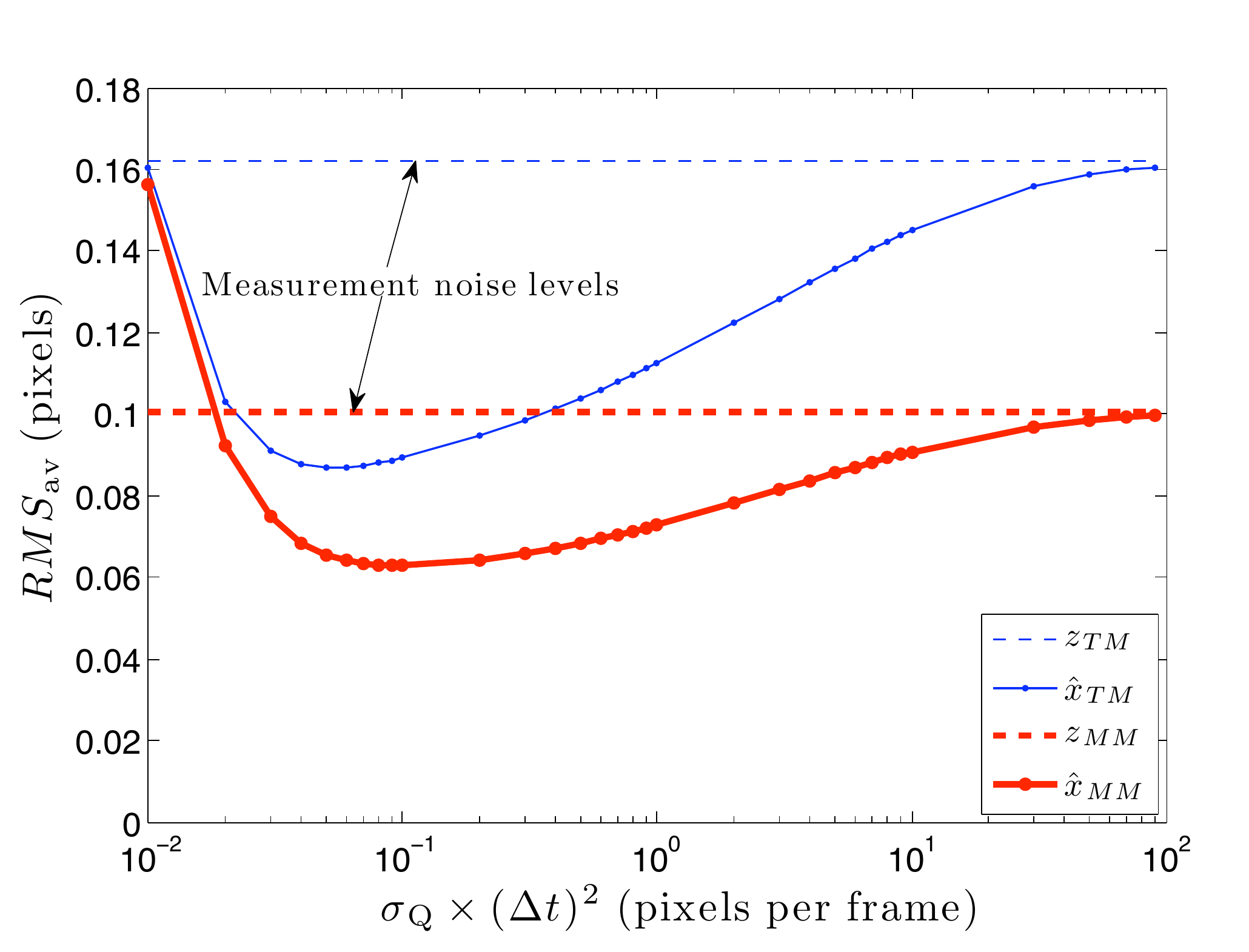}
      	    \caption{\label{fig:PNtune}
      	      For a single ensemble of 1000 virtual (simulated) experiments, multiple EKFs were run using 
      	      a different process noise magnitude, expressed here in units of pixels per frame: 
      	      $\sigma_\mathrm{Q} \times (\Delta{t})^2$.  The optimal values (corresponding to fluctuations in 
      	      particle positions of one-tenth and one-twentieth of a pixel per frame for 
      	      MM and TM measurements, respectively) minimize the average RMS error (given here in pixels) 
      	      over long times.}
          }
        \end{figure}

    	  A tunable parameter for any EKF (or simple KF) is the process noise magnitude 
    	  $\sigma_\mathrm{Q}$, which models the standard deviation for the highest-order 
    	  time-derivative of position that is not included in the prediction model -- 
    	  in this case, the \emph{jerk} (the time-derivative of acceleration).  
    	  As these typical fluctuations are generally unknown a priori, $\sigma_\mathrm{Q}$ 
    	  becomes a design parameter that needs to be optimized to help minimize modeling errors.  
    	  This ``process noise tuning'' involves minimizing the average mean-square or 
    	  root-mean-square (RMS) error as a function of $\sigma_\mathrm{Q}$.  
    	  This is one example of where virtual experiments are beneficial because many 
    	  experiments (perhaps hundreds or thousands) can be simulated to obtain good 
    	  averages.  
    	  For a typical simulation ensemble of size $N = 1000$, $\sigma_\mathrm{Q}$ was varied over 
    	  a finite range and the resulting average RMS error in the state estimate of the 
    	  target particle's position, $(q_\mathrm{x},q_\mathrm{y})$ was calculated as\cite{BarShalom}
    	  \begin{equation}
    	    \label{eq:RMS}
    	    RMS(\sigma_\mathrm{Q},k) = \sqrt{\frac{1}{N}\sum_{i=1}^N \sq{(\Delta q_{\mathrm{x},i}(k))^2 + (\Delta q_{\mathrm{y},i}(k))^2}},
    	  \end{equation}
    	  where $\Delta q_{\mathrm{x},i}(k)$ is the deviation of the estimated x-position from the true x-position at 
    	  time $t(k)$ (and similarly for the y-direction).  The long-time average of (\ref{eq:RMS}) provides a convenient 
    	  measure of the filter's performance as a function of $\sigma_\mathrm{Q}$.  This is shown in 
    	  \Fig{fig:PNtune} (note the logarithmic horizontal axis), where data corresponding to TM measurements 
    	  is blue and MM measurements is red (and bold).  
    	  For increasingly large process noise magnitude, confidence in the prediction model decreases so that 
     	  the measurements are ``trusted'' more by the filter and the RMS error in the estimated state 
     	  asymptotically approaches the measurement noise level.  
    	  An optimal process noise magnitude yields a minimum RMS error in the estimated state -- note that 
    	  the TM state estimate for a tuned filter performs better than the (non-optimized) MM measurement 
    	  level, providing very encouraging support for a software-based approach to improving 
    	  particle-locating precision, as opposed to what could be a laborious hardware optimization 
    	  process in a real experiment in order to obtain comparable errors sizes.  
    	  In any case, the state estimate outperforms the measurements for sufficiently large process noise, 
    	  showing that it's almost always beneficial to use \emph{any} knowledge you have of the target 
    	  particle dynamics to improve track precision via an EKF (the payoff decreases exponentially for 
    	  less certain knowledge as the information processing costs outweigh the reduced improvement in 
    	  precision).  
    	  
    	  The optimal values of process noise magnitude were 
    	  $\sigma_\mathrm{Q,TM}(\Delta{t})^2 = 0.05$ pixels per frame and 
        $\sigma_\mathrm{Q,MM}(\Delta{t})^2 = 0.1$ pixels per frame, where the sampling period used 
        corresponds to imaging the virtual experiment at 200 frames per second: $\Delta{t}=0.005s$.  
        The lower precision of TM measurements should result in a higher level of confidence 
        (smaller $\sigma_\mathrm{Q}$) in the prediction model, and so the RMS errors in the 
        state estimate are minimized for $\sigma_\mathrm{Q,TM} < \sigma_\mathrm{Q,MM}$.

  	  \subsection{Filter consistency}
  	    For static parameter estimation, consistency of an estimator is defined as convergence 
  	    of the estimate $\hat{x}$ to the true value $x$.  
  	    For dynamic parameter estimation, estimator consistency is defined in terms of 
  	    conditions on the first two moments of the estimate:\cite{BarShalom}
  	    \begin{subequations}
    	    \begin{eqnarray}
    	      E[x - \hat{x}] & = &0 ,\label{eq:consistency1} \\
    	      E[\cu{x = \hat{x}}\cu{x = \hat{x}}^\prime] &=& P .\label{eq:consistency2} 
    	    \end{eqnarray}
  	      \label{eqs:consistency}\end{subequations}
  	    The first is satisfied by an unbiased estimator (zero-mean estimation error).  
  	    The second condition is that of covariance matching between the MSE of the estimates 
  	    (left-hand side) and the filter-calculated covariance (right-hand side).  
  	    The two conditions in \Eq{eqs:consistency} can be tested simultaneously using a 
  	    $\chi$-squared test, as described in many undergraduate-level statistics texts -- 
  	    Ref.~\citenum{BarShalom} describes it in the context of a Kalman filter.  
  	    When the test statistics reside within an acceptance window, then the filter is 
  	    determined to be consistent, which was the case in this work 
  	    (although they are not presented here).

  	%%%%%%%%%%%%%%%%%%%%%%%%%%%%%%%%%%%%%%%%%%%%%%%%%%%%%%%%%%%%%
    \section{RESULTS AND DISCUSSION}\label{sec:results}
      \begin{figure}[htp]
        {\centering 
        \subfigure[~Sample particle trajectory\label{fig:trajectory}]{\includegraphics[width=0.49\columnwidth]{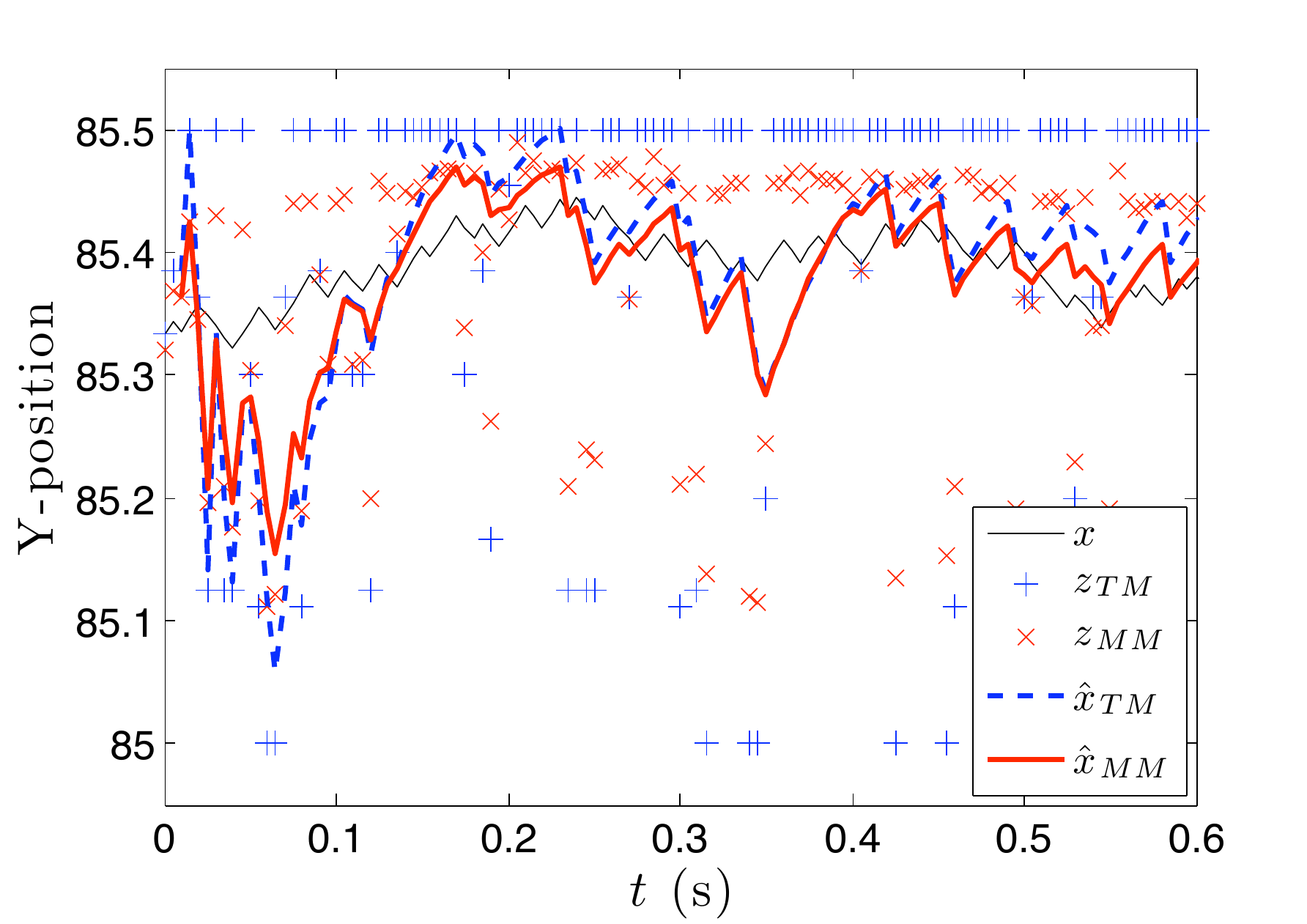}}  
        \subfigure[~Optimal RMS errors over time\label{fig:RMSt}]{\includegraphics[width=0.49\columnwidth]{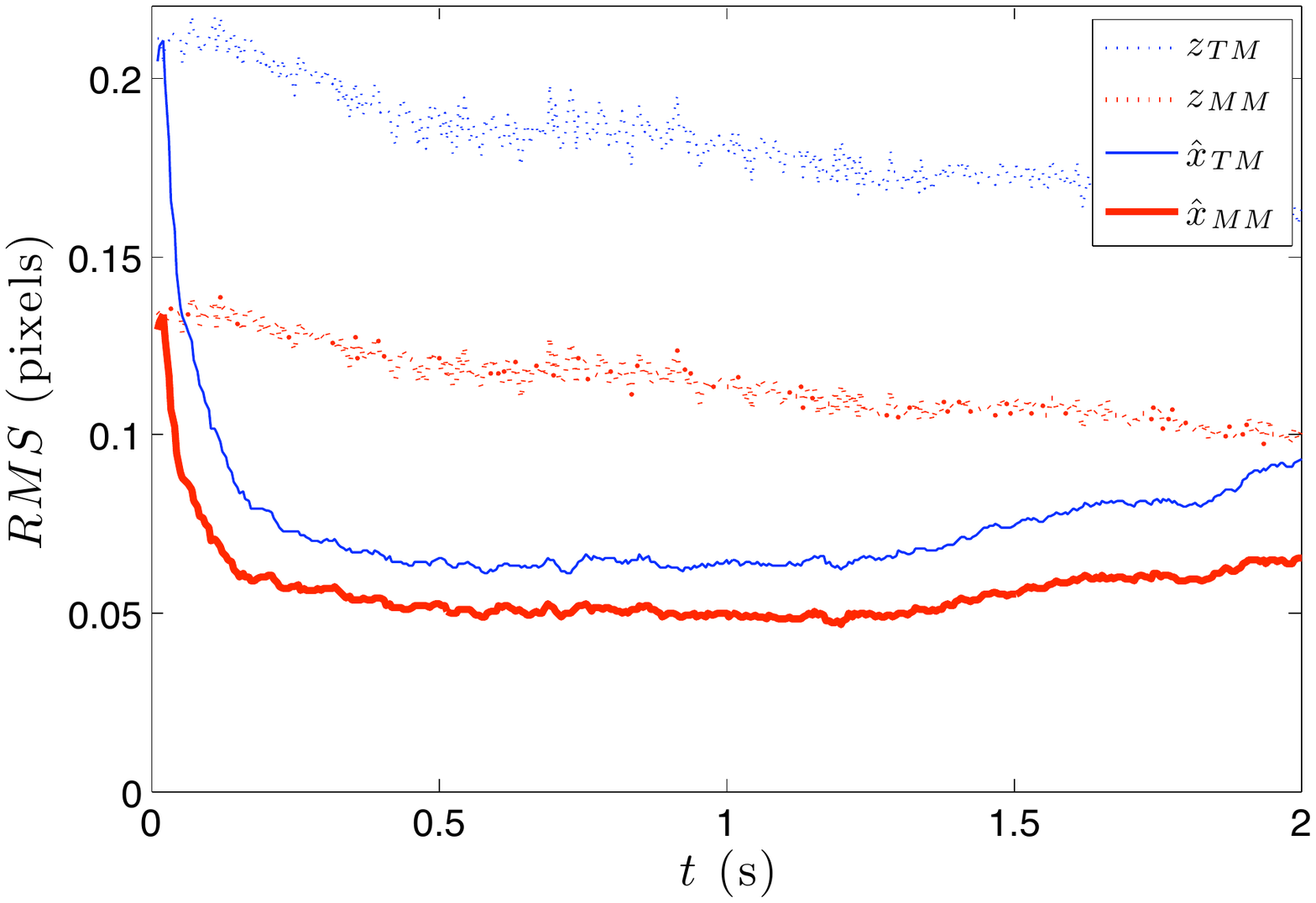}} 
          \caption{\label{fig:Results}
            Shown in \subref{fig:trajectory} is a sample trajectory showing the true evolution of the target's Y-position 
            (thin black line), a spread of TM (blue $+$) and MM (red $\times$) measurements and the 
            corresponding state estimates for the tuned process noise values given in \Fig{fig:PNtune}.  
            Note the significantly larger deviation of the measurements compared to the estimates.  
            Shown in \subref{fig:RMSt} are the RMS errors in measured and estimated positions as a function of time, 
            again for the tuned values of process noise.
            The sample period was $\Delta{t}=0.005$s (200 frames per second).
            }
        }
      \end{figure}
      \Fig{fig:trajectory} is a sample trajectory from the early stages of a single virtual experiment.  
      The true Y-position of the central particle is shown as a thin black line, with the MM and TM measurements 
      shown as crosses ($\times$) and plus signs ($+$), respectively.  Shown are the state estimates generated by 
      the EKF after processing each measurement and weighing it with the prediction model -- the red line (MM) 
      and blue dashed line (TM).  
      Note the transient period before the filter ``settles down'' and tracks the true state quite well.
      
      \Fig{fig:RMSt} shows the RMS errors for the measurements (dotted lines) and estimated positions (solid lines) 
      of the target particle when tracking only the target, but feeding the NN measurements into the prediction 
      stage of the EKF.  When tracking all 7 particles in the hexagonal ``crystal'' cell, the increased precision 
      in the tracked NN positions did not significantly improve the target particle track.  This is reasonable 
      because the measurements were quite good, as shown earlier.  For situations where the measurements are not so 
      good (due to noisier images, for example), or the NN particles are moving faster, tracking the NNs (as opposed 
      to just measuring them{, which can represent a massive computational saving when tracking many particles}{}) 
      is expected to produce noticeably better results for the target track{}{, albeit at a noticeably higher 
      computational expense due to the increased number of states in the filter.}
      For longer times than shown, the RMS error in the estimated position exhibits low-magnitude oscillations at 
      a rate corresponding to the rate of oscillation of the interacting dust particles about the initial positions.  
      This explains the apparent slight increase in the state estimate RMS errors at later times in \Fig{fig:RMSt}.
      
      As shown in \Sec{sec:PNtune}, the process noise magnitude $\sigma_\mathrm{Q}$ can be tuned to improve the EKF 
      performance (minimize the errors in the estimated state).  In the same way that the magnitude 
      $\sigma_\mathrm{Q}$ represents confidence in the prediction model, the measurement noise magnitude 
      $\sigma_\mathrm{R}$ represents confidence in the measurements.  
      \Fig{fig:PNtune} shows that the actual measurement errors were quite small: 
      $\sigma_\mathrm{R,TM} \approx 0.16$ pixels, and $\sigma_\mathrm{R,MM} \approx 0.1$ pixels.  In this work, no knowledge 
      of the measurement precision was assumed and the so-called\cite{hadziavdic2006} ``worst-case scenario'' 
      of $\sigma_\mathrm{R}^2 = 0.1$ was used.  Overestimating the measurement noise in this way is usually 
      not as detrimental to the resulting performance of a Kalman filter as underestimating it, as shown for 
      a simplified Kalman filter in Ref.~\citenum{hadziavdic2006}.  Therefore, it is expected that using a 
      measurement noise size in the EKF that is closer to the true value would not significantly improve the 
      EKF performance.  
      Note that a similar tuning process can be performed for the measurement noise as was done for the 
      process noise in \Sec{sec:PNtune}. 
      
      \begin{figure}[htp]
        {\centering 
          \subfigure[~True position\label{fig:subpixel:true}]{\includegraphics[width=0.29\columnwidth]{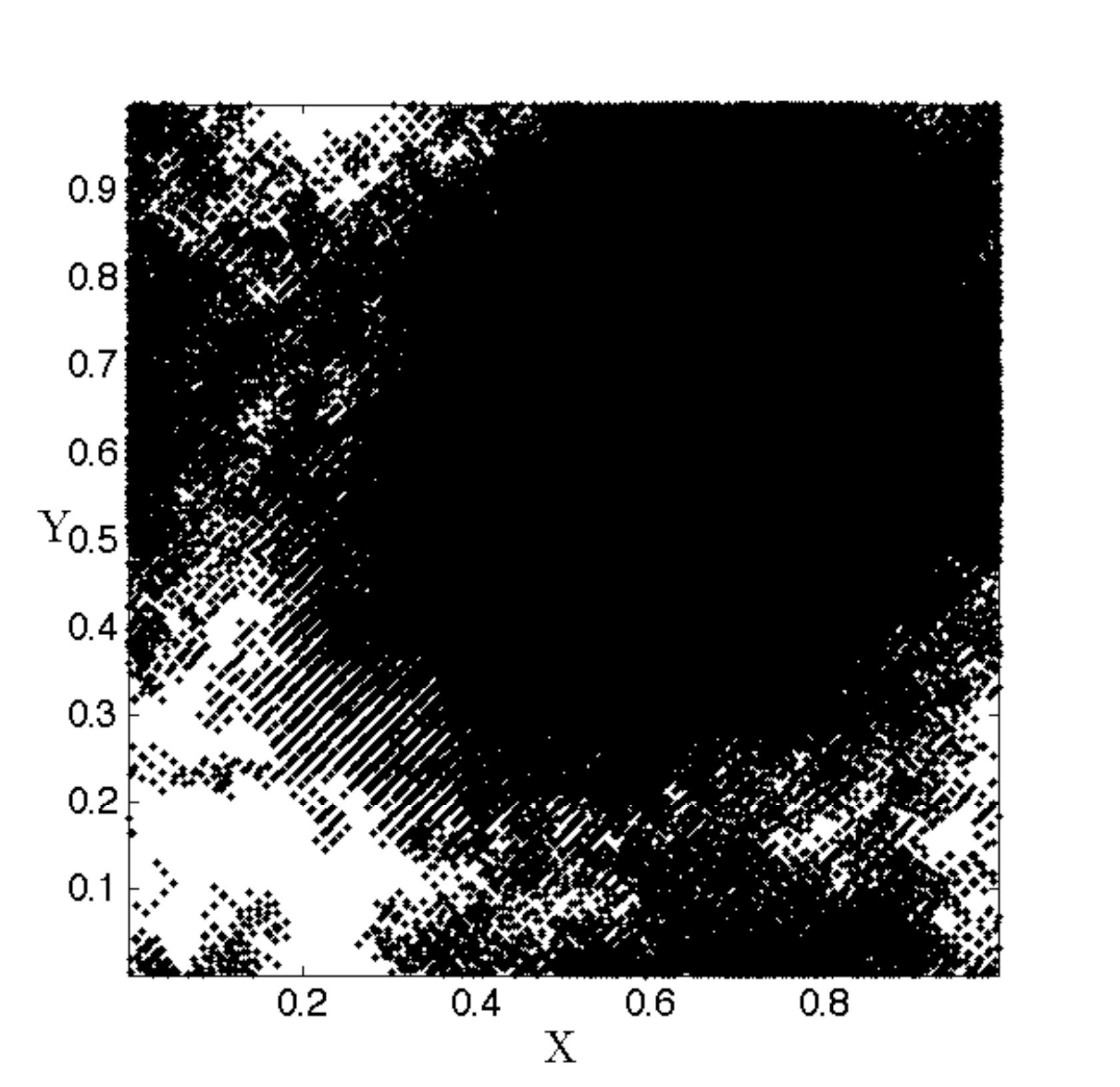}}
          \subfigure[~TM-measured position\label{fig:subpixel:TM}]{\includegraphics[width=0.29\columnwidth]{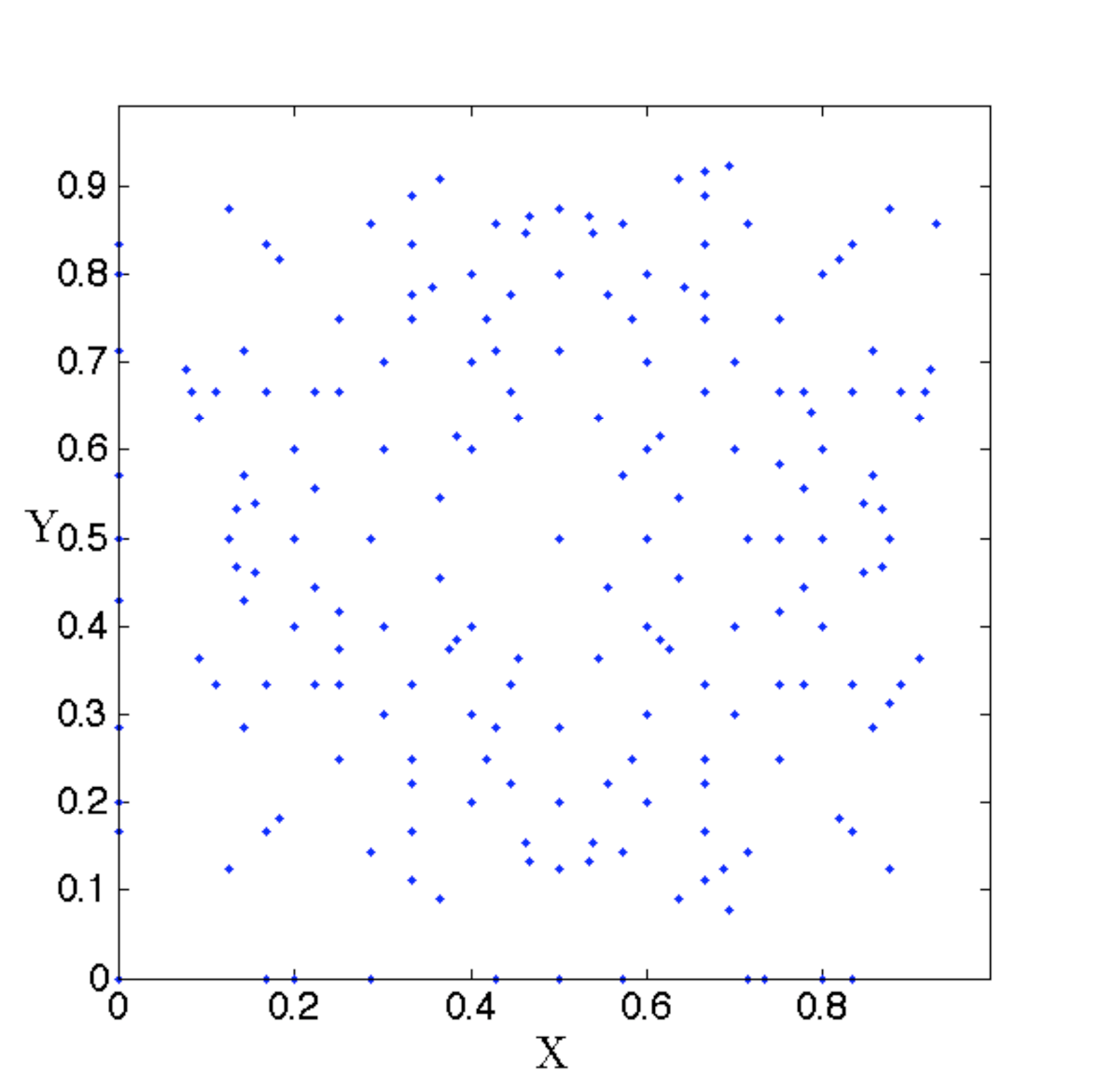}}
          \subfigure[~MM-measured position\label{fig:subpixel:MM}]{\includegraphics[width=0.29\columnwidth]{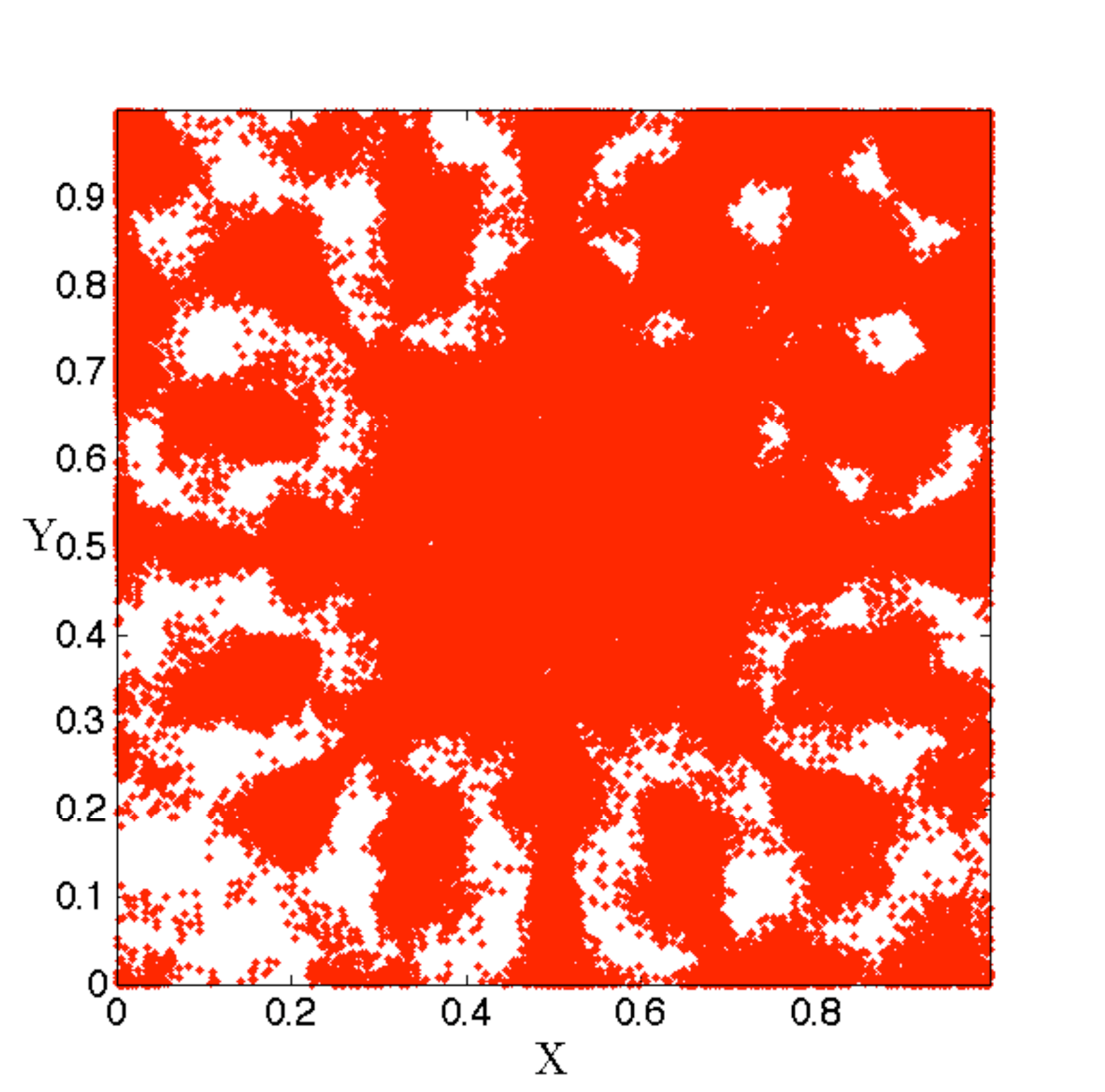}}\\ 
          \subfigure[~TM-based estimated position\label{fig:subpixel:TMest}]{\includegraphics[width=0.29\columnwidth]{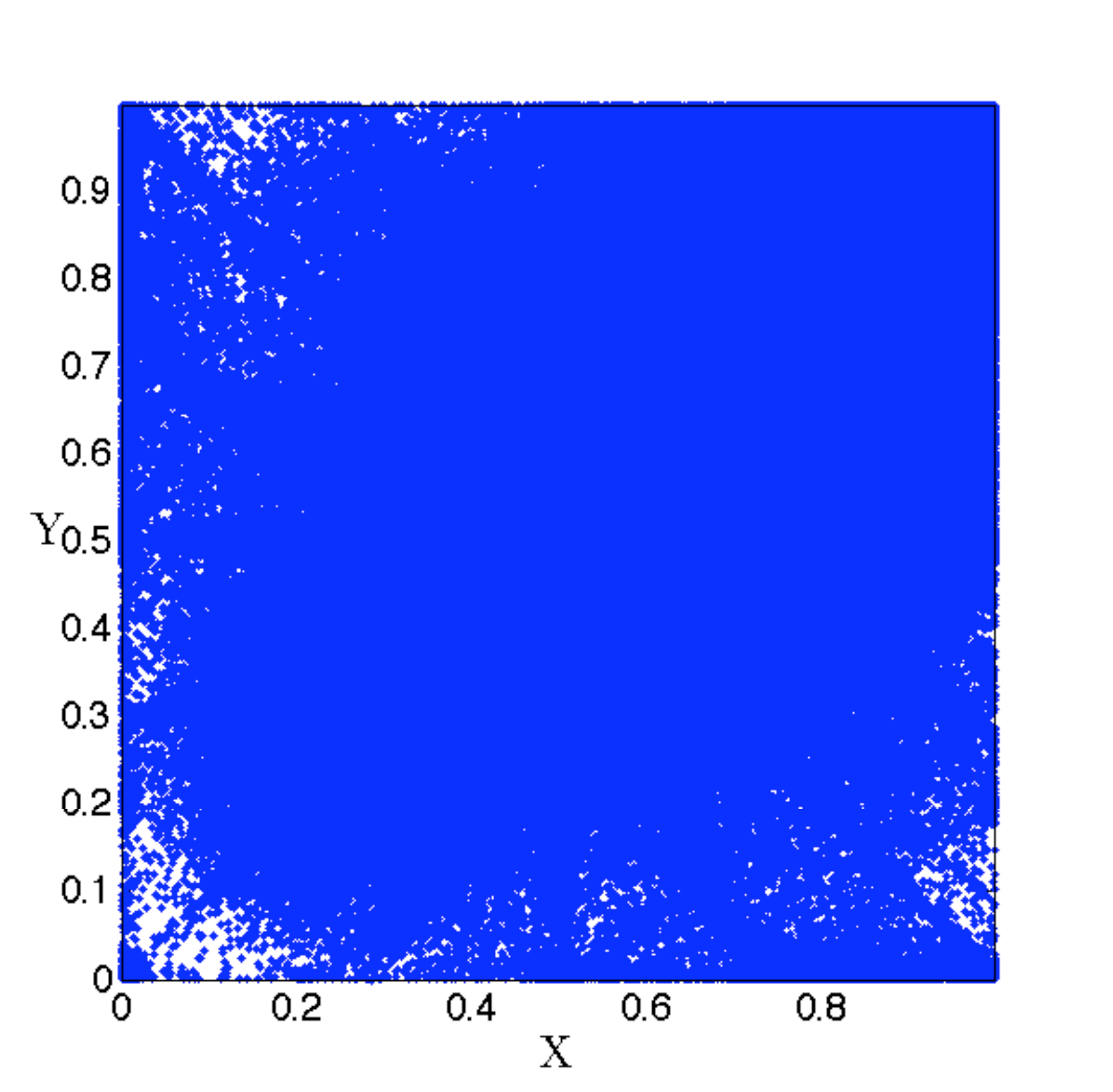}}
          \subfigure[~MM-based estimated position\label{fig:subpixel:MMest}]{\includegraphics[width=0.29\columnwidth]{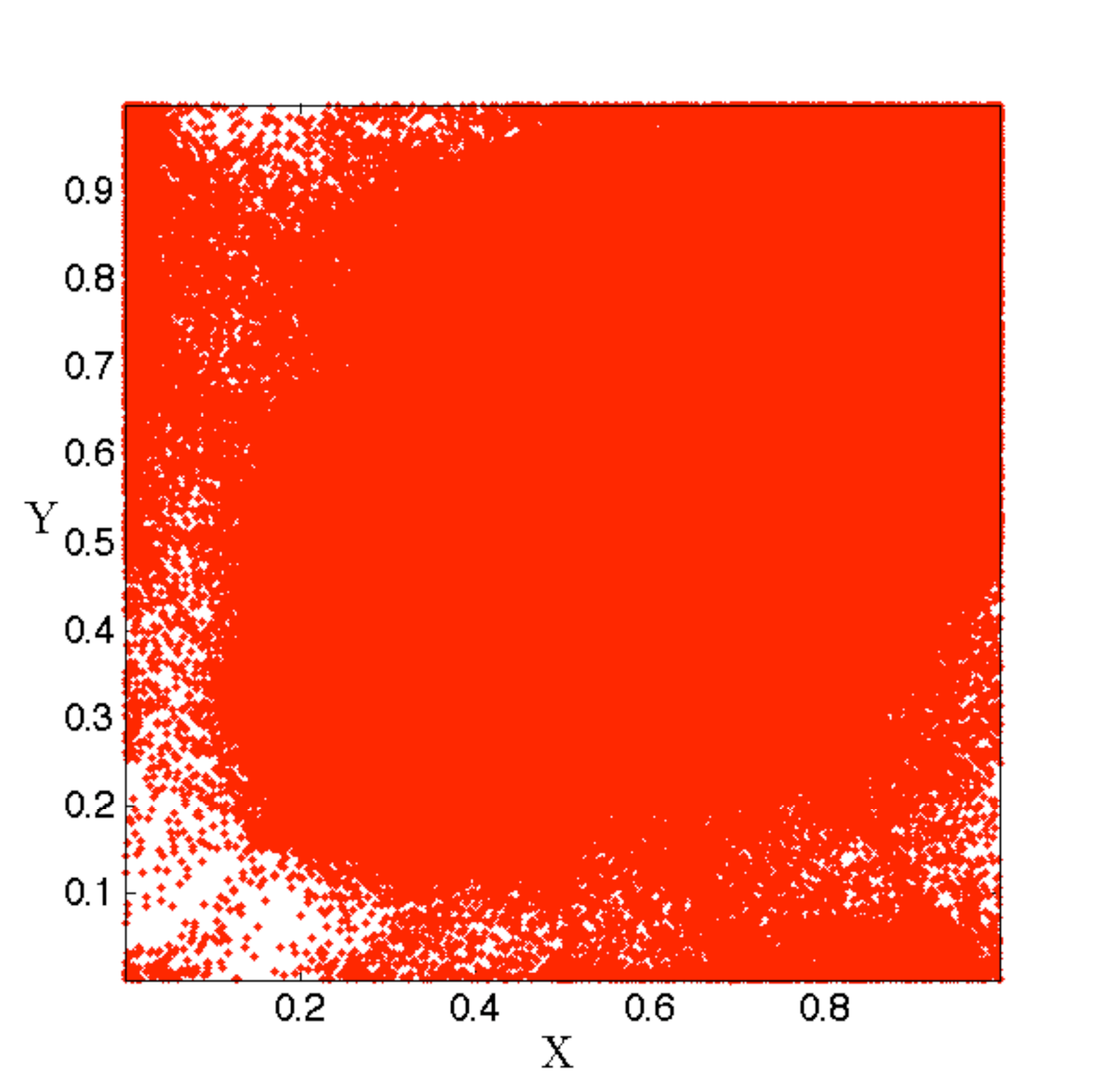}}  
            \caption{\label{fig:subpixel}
              Sub-pixel maps for the central particle's 
              \subref{fig:subpixel:true} true positions, 
              \subref{fig:subpixel:TM} TM-measured positions, 
              \subref{fig:subpixel:MM} MM-measured positions, 
              \subref{fig:subpixel:TMest} TM-based estimated positions, 
              \subref{fig:subpixel:MMest} MM-based estimated positions, across 1000 virtual experiments.  
              Note the presence of pixel-locking in the TM measurements in \subref{fig:subpixel:TM} 
              (and to a lesser extent in the MM measurements in \subref{fig:subpixel:MM}), 
              and its subsequent removal by the EKF in \subref{fig:subpixel:TMest}.
              }
        }
      \end{figure}
      There are a few notable potential sources of error in tracking particles in dusty plasma 
      experiments (both virtual and real).  They can be categorized as measurement-based, model-based and 
      numerical.  The latter come in the form of round-off errors that can be minimized by using the 
      Joseph form of the covariance update equation in the EKF.\cite{BarShalom} 
      Modeling errors, where an incorrect model is used for the system dynamics, can lead to incorrect 
      predictions, and over- or under-confidence in predictions (versus measurements) due to sub-optimal 
      process noise values -- overcome here by the process-noise-tuning process.  
      The most serious measurement-based errors for the rather unsophisticated techniques considered here 
      (TM and MM) come in the form of ``pixel-locking'',\cite{Feng2007,Ivanov2007} resulting from 
      insufficient pixel-resolution for the particles.  This results in certain subpixel regions being 
      preferred, as shown in \Fig{fig:subpixel} where the measured subpixel locations (\Fig{fig:subpixel:TM} 
      and \subref{fig:subpixel:MM}) of the target particle 
      in 401 images from each of 1000 virtual experiments (401,000 data points) are plotted 
      alongside the true (\Fig{fig:subpixel:true}) and estimated (\Fig{fig:subpixel:TMest} and 
      \subref{fig:subpixel:MMest}) subpixel locations.  
      Pixel-locking was reduced for the MM measurements (see \Fig{fig:subpixel:MM}) in a fairly standard way 
      -- ``smearing'' the particles slightly by defocussing the virtual camera.  
      Remarkably, although this technique was completely unsuccessful for the TM measurements 
      (see \Fig{fig:subpixel:TM}), the filter almost completely removed the pixel-locking effect as 
      shown in the subpixel map for the TM-based estimates in \Fig{fig:subpixel:TMest}.  
      These maps were representative of all the particles.  
      They show that strong pixel-locking occurred for the TM measurements (and to a lesser extent for the 
      MM measurements), but was removed after processing these measurements through the filter.

    %%%%%%%%%%%%%%%%%%%%%%%%%%%%%%%%%%%%%%%%%%%%%%%%%%%%%%%%%%%%%
    \section{CONCLUSION}\label{sec:conclusions}
      An extended Kalman filter (EKF) algorithm has been designed and implemented for tracking 
      interacting dusty plasma particles.  {The algorithm was tested in virtual 
      experiments (simulations) which allowed for quantification of performance aspects such as 
      the precision of locating particle positions}.  
      Attention was paid to generating images {that closely} resemble real experiments.  
      The algorithm {was tested for a crystal-like 
      phase of the dusty plasma.  This simple scenario provides a benchmark for measuring 
      the performance of more sophisticated algorithms that will be necessary for treating more 
      complex dynamics.}  
      An analytical result showed the validity of an approximation to only nearest-neighbor 
      interactions for this situation.  
      Starting with relatively imprecise measurements of particle positions using the 
      threshold and moment methods, it was shown that the filter algorithm improved the 
      precision, as well as removing or significantly reducing errors due to ``pixel-locking''.  
      The RMS values of position errors exceeded $0.1$ pixels for the measurements, which were 
      improved by the filter to approximately $0.06$ pixels -- approaching the precision obtained in 
      Ref.~\citenum{Ivanov2007} ($\approx0.03$ pixels) where a much more sophisticated and 
      time-consuming measurement technique was used (local quadratic kernel method).  
      This suggests that when such ``cheap'' measurements are used in conjunction 
      with an EKF, they may be sufficient for obtaining excellent sub-pixel precision for particle 
      positions -- a task that was previously {achieved through hardware modifications and/or} 
      the aforementioned sophisticated measurement (image-processing) techniques.  
      This particular result bodes well for eventually applying such a particle-tracking filter in 
      real-time, {perhaps as part of a feedback loop.}
      
      It is expected that the algorithm's {will perform quite well}{performance} in 
      experimental situations {}{will be comparable to the results presented here}, 
      but performance will degrade the further a dusty plasma deviates from a crystal-like state.  
      Future work includes applying the algorithm to experimental images, 
      extending the algorithm to cope with highly nonlinear dynamics such as shock-waves and 
      Mach cones, and considering more advanced state estimation and measurement-track association 
      techniques.  The present results will provide a useful performance benchmark for future work.

%%%%%%%%%%%%%%%%%%%%%%%%%%%%%%%%%%%%%%%%%%%%%%%%%%%%%%%%%%%%%
\acknowledgments
  
  The authors acknowledge financial support from UK EPSRC grant number EP/G007918.  
  N.P.O.~acknowledges use of high-throughput computational resources provided by the 
  eScience team at the University of Liverpool.

%%%%%%%%%%%%%%%%%%%%%%%%%%%%%%%%%%%%%%%%%%%%%%%%%%%%%%%%%%%%%
%%%%% References %%%%%
\bibliographystyle{spiebib}
\bibliography{}

\end{document}